\documentclass[preprint,12pt,<options>]{elsarticle}

\usepackage[utf8]{inputenc}
\usepackage{amssymb}
\usepackage{xcolor}
\usepackage{amsmath}
\biboptions{sort&compress}

\begin{document}

\begin{frontmatter}



\title{Anomalous temperature dependence of the electrical resistivity in R$_3$Co$_4$Ge$_{13}$ (R = Y, Lu) single crystals}


\author{Juliana Gonçalves Dias}
\author{Shyam Sundar\corref{cor1}\fnref{label2}}
\ead{shyam.phy@gmail.com}
\author{Leticie Mendonça-Ferreira} 
\author{Marcos A. Avila\corref{cor1}}
\ead{avila@ufabc.edu.br}
\affiliation{organization={Centro de Ciências Naturais e Humanas (CCNH), Universidade Federal do ABC}, 
             city={Santo André},
             postcode={09210-580},
             state={SP},
             country={Brazil}}
\cortext[cor1]{Corresponding authors}
\fntext[fn1]{Presently at: Instituto de Fisica, Universidade Federal do Rio de Janeiro, 21941-972 Rio de Janeiro, RJ, Brazil.}

\begin{abstract}
The presence of strong disorder can significantly impact electrical conduction in metallic systems. 
Here, we investigate the temperature dependence of the electrical resistivity, $\rho(T)$, in nonmagnetic single crystals of the Remeika-phase cage compounds R$_3$Co$_4$Ge$_{13}$ (R = Y, Lu). 
Contrary to the density of states (DOS) calculations in the literature, the experimentally measured $\rho(T)$ in both compounds exhibits semiconducting-like behavior, which we attribute to the strong structural disorder due to its unique crystal structure and low carrier-density. 
A detailed analysis of the electrical resistivity data reveals that neither the Arrhenius thermal activation law nor variable-range hopping (VRH) models can adequately describe their temperature dependence over the broad temperature range of 2-350 K.
However, a model incorporating parallel conduction through both semiconducting and metallic channels provides an adequate explanation. 
In addition to a dominant metallic conduction below $\sim 10$~K, a negative temperature coefficient of the electrical resistivity ($d\rho/dT$) is found in both samples. 
In the absence of magnetic impurities, the observed $d\rho/dT < 0$ is interpreted in terms of the structural Kondo mechanism.

\end{abstract}

\begin{graphicalabstract}
\centering
\includegraphics[width=1\textwidth]{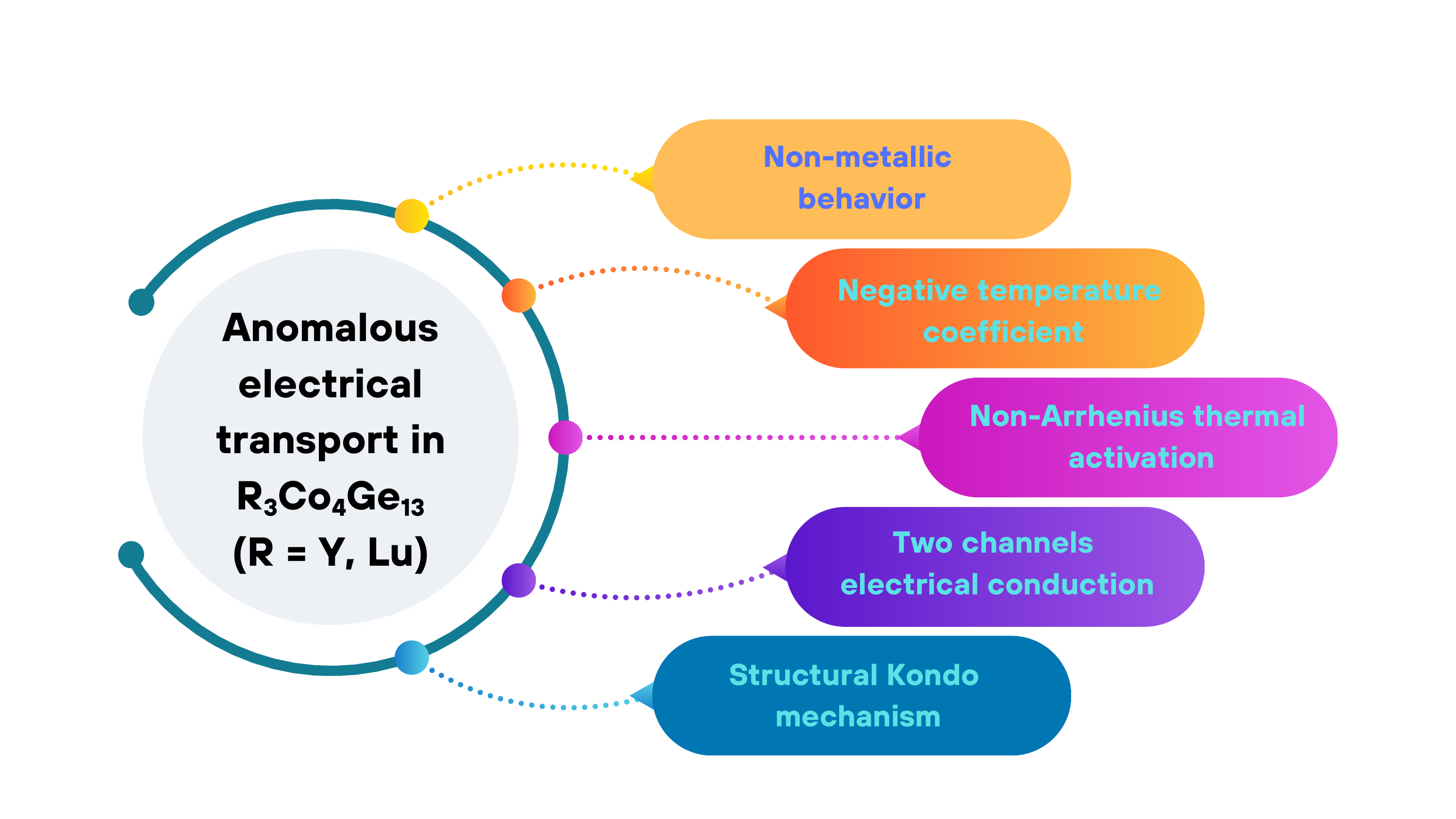}
\end{graphicalabstract}

\begin{highlights}
\item We report the structural and electrical transport properties of the R$_3$Co$_4$Ge$_{13}$ (R = Y, Lu) single crystals.
\item Temperature dependence of the electrical resistivity does not follow the conventional transport mechanisms in both compounds.  
\item Parallel conduction, consisting of semiconducting and metallic channels, reasonably explains the electrical resistivity data.
\item Structural disorder plays a crucial role in electrical transport, in terms of the structural Kondo mechanism, in both compounds. 

\end{highlights}


\end{frontmatter}

\section{Introduction}
\label{sec1}

The Remeika phase consists of intermetallic ternary compounds with stoichiometry R$_3$T$_4$X$_{13}$ (R - Rare-earth; T - Transition metal; X = Ge, Sn and In), also known as 3-4-13 compounds. These materials belong to the family of intermetallic cage-compounds, where a positively charged filler atom located inside an oversized cage formed by a network of negatively charged covalent bonds \cite{remeika1980new, GUMENIUK201843, oswald2017proof}. Remeika phase compounds display a wide variety of interesting physical phenomenon including unconventional superconductivity, magnetism, Kondo effect and heavy-Fermions, as well as topological insulating behaviour \cite{feig2021valence, rai2016intermediate, prakash2013superconductivity, rai2015superconductivity, kataria2023time, kamadurai2021semiconducting}. 
Another class of intermetallic cage compounds, the filled skutterudites, exhibits similar physical behavior but is structurally distinct from the Remeika phases. These materials have been identified as technologically relevant for thermoelectric (TE) applications, where the dimensionless thermoelectric figure of merit ($ZT$) quantifies their performance \cite{ZHANG2020152272, UHER2001139}. Therefore, there has been some interest to explore the thermoelectric properties of Remeika phases compounds as well. However, the known Remeika phase compounds show significantly lower thermoelectric performance as compared to the filled skutterudites \cite{GUMENIUK201843, UHER2001139}. Optimization of $ZT$ in a thermoelectric material relies on a large thermopower $S$ as the key ingredient, but it also depends on the intricate relation between electrical and thermal conduction ($\sigma/\kappa$). Due to their unique crystal structures, the Remeika compounds tend to possess significant structural defects, which make their electrical and thermal transport mechanisms quite complex, but also provide materials engineering routes towards optimization of $ZT$. Therefore, it is important to understand and control the transport properties in these compounds to enhance their TE performance \cite{GUMENIUK201843, nolas1999skutterudites, rull2015skutterudites,MOROZKIN2013121, Ogunbunmi_2020}.

In contrast to the stannide compounds in the Remeika family which show metallic temperature dependence of the electrical resistivity, their germanide counterparts exhibit narrow-gap semiconducting or semimetallic character below 300 K \cite{rai2015superconductivity}. 
Irrespective of the metallic or non-metallic resistivity in the normal state, non-magnetic stannide/germanide-based Remeika compounds are known to present unconventional superconductivity at very low temperatures \cite{rai2015superconductivity, prakash2013superconductivity, kataria2023time}. 
Such manifestation of superconductivity in these compounds is particularly interesting due to their low-carrier density \cite{rai2015superconductivity} and non-BCS characteristics \cite{prakash2014superconductivity, kataria2023time}. 
Despite the existing reports on the investigation of unconventional superconductivity in Remeika compounds, no study has so far aimed to understand the unusual electrical transport behavior in the semiconducting normal state of the Ge-based compounds, which should be a determining factor enabling unconventional superconductivity.
In contrast to the experimental observations, density functional theory (DFT) calculations show that the Fermi level in 3-4-13 germanides is close to a local maximum, indicative of metallic behavior \cite{rai2015superconductivity}. 

It is well known that, in the absence of relevant electronic correlations, strong disorder may lead to electron localization and can change a metal into an insulator through Anderson localization \cite{mott1970conduction}. 
Rai \textit{et al.} \cite{rai2015superconductivity} suggested that the experimental semiconductor-like behaviour in 
3-4-13 germanides is most likely due to the large amount of disorder at the structural Ge1 site. 
Therefore, a systematic study of the temperature dependence of electrical resistivity is in demand.
Here, we report a detailed investigation on the electrical transport of Y$_3$Co$_4$Ge$_{13}$ and Lu$_3$Co$_4$Ge$_{13}$ single crystals, which are non-magnetic in nature and exhibit a semiconducting-like behavior between 2-350~K. 

We show that the temperature dependence of electrical resistivity $\rho(T)$ for both compounds does not follow the usual thermal activation behaviour of standard semiconductors, nor can be described by a common variable range hopping transport mechanism. 
Rather, $\rho(T)$ in the entire temperature range is reasonably explained in both compounds by considering the combination of semiconducting and metallic conduction channels where, at low-temperatures, metallic conduction is found to be more effective.
In addition to a negative temperature coefficient of resistivity ($d\rho/dT<0$) in a wide temperature range below 350 K, a log-log plot of $\rho(T)$ demonstrates gradual leveling-off of resistivity at the lowest temperatures. 
Such behaviour can be described in terms of structural Kondo type scattering mechanism, which involves an attractive interaction between conduction electrons and localized excitations arising from crystallographic site disorder.

\section{Experimental Methods}
\label{sec2}

Single crystals of Y$_3$Co$_4$Ge$_{13}$ and Lu$_3$Co$_4$Ge$_{13}$ were grown by the self-flux method, 
using a molar ratio of 10:15:75 (R:Co:Ge) for the initial mixture. 
Elemental reagents were supplied by Alfa Aesar and Sigma-Aldrich with purities of 99.9999$\%$ for Ge, and 99.9$\%$ for Co, Y and Lu elements.
The elements were loaded in quartz ampoules, sealed under vacuum, and placed in a box furnace.
The ampoules were heated from room temperature up to 1100~$^{\circ}$C, and maintained at that temperature for 2~h, during which all elements of the mixture were melted into a homogenous liquid. 
Next, the ampoules were cooled to 1000~$^{\circ}$C over 4~h and then slowly cooled to 900~$^{\circ}$C over 100~h. 
At this temperature, the ampoules were removed from the furnace and centrifuged to separate the crystals from the molten flux \cite{dias2025single}.

For both samples, the cubic crystal structure consistent with the Remeika phase was confirmed by powder XRD measurements at room temperature on a Stoe STADI P diffractometer with a Ge (111) monochromator and Debye-Scherrer geometry, using Mo K$\alpha_1$ radiation. 
Rietveld refinement analysis of the powder pattern was performed using TOPAS-Academic v7 (TA-7) \cite{coelho2018topas}. 
More information on the details of the crystal structure of these compounds may be found in Ref.~\cite{dias2025single} and references therein. 
Sample homogeneity and elemental composition were confirmed using EDXS coupled to a JEOL JSM-6010LA Scanning Electron Microscopy (SEM). 
Electrical resistivity was measured in a Quantum Design Physical Property Measurement System (QD-PPMS) using a standard 4-probe technique in the temperature range 2-350~K.
Crystals of both compounds were cut into rectangular slabs and then polished before attaching four Pt-wires, for current and voltage leads, using Epotek H20E silver epoxy. 
Electrical contacts were cured at 400~K in a box furnace before mounting the samples on the sample holder for low-temperature measurements.

\begin{figure}[h]
 \centering
\includegraphics[width=0.7\textwidth]{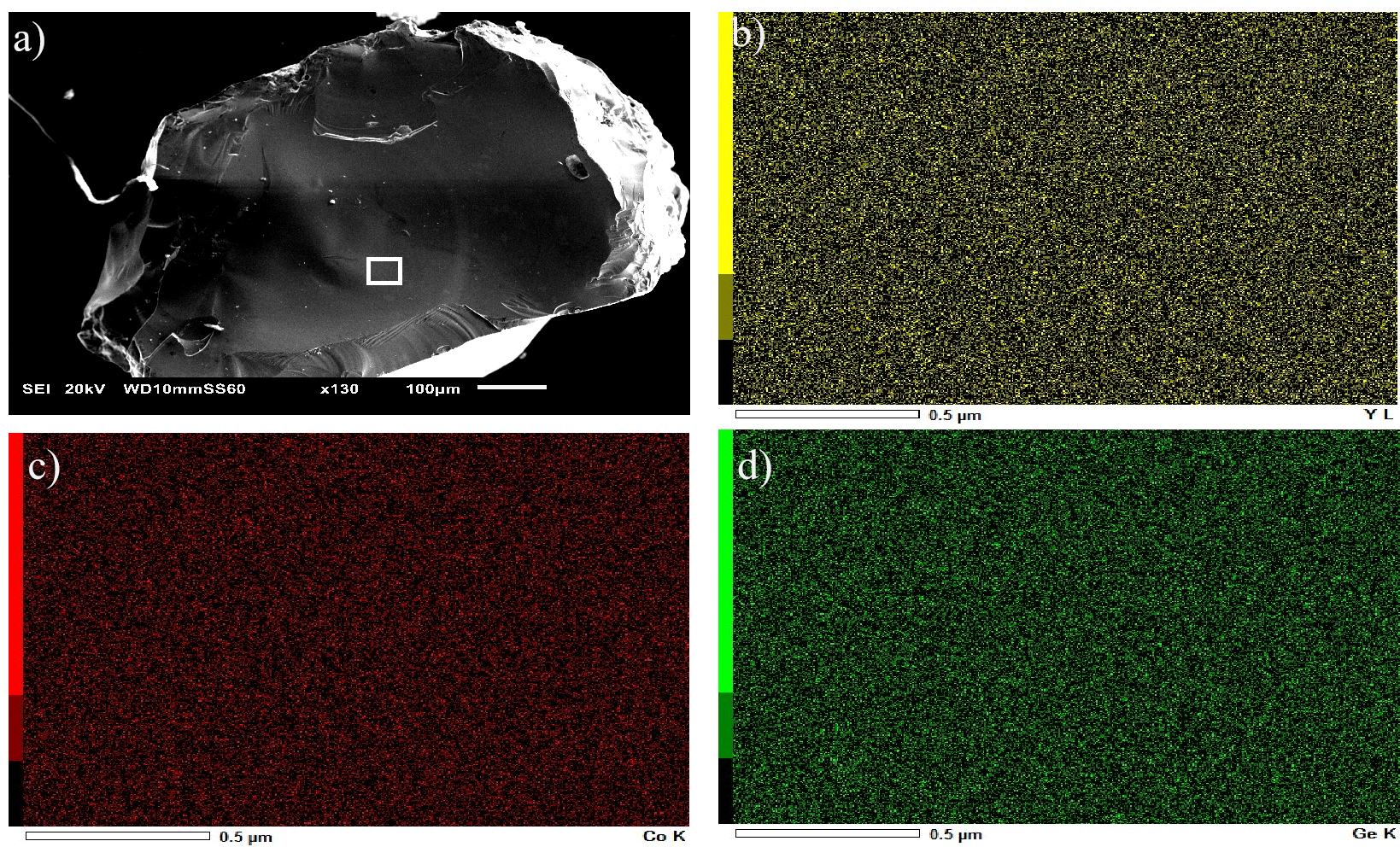}
\caption{(a) SEM image of an Y$_3$Co$_4$Ge$_{13}$ crystal. The white rectangle indicates the selected area for the EDX map. Individual elemental analysis for (b) Y, (c) Co, and (d) Ge.}
\label{fig:fig8}
\end{figure}

\section{Results and Discussion}
\label{sec3}

The average effective stoichiometry of the Y$_3$Co$_4$Ge$_{13}$ and Lu$_3$Co$_4$Ge$_{13}$ crystals estimated from the EDXS measurements (Fig.~\ref{fig:fig8}) is Y$_{3.4(8)}$Co$_{4.4(7)}$Ge$_{12.0(2)}$ and Lu$_{3.0(4)}$Co$_{4.2(2)}$Ge$_{12.6(5)}$, consistent with the Remeika phase within the measured precision.

\begin{figure}[htpb]
 \centering
\includegraphics[width=0.4\textwidth]{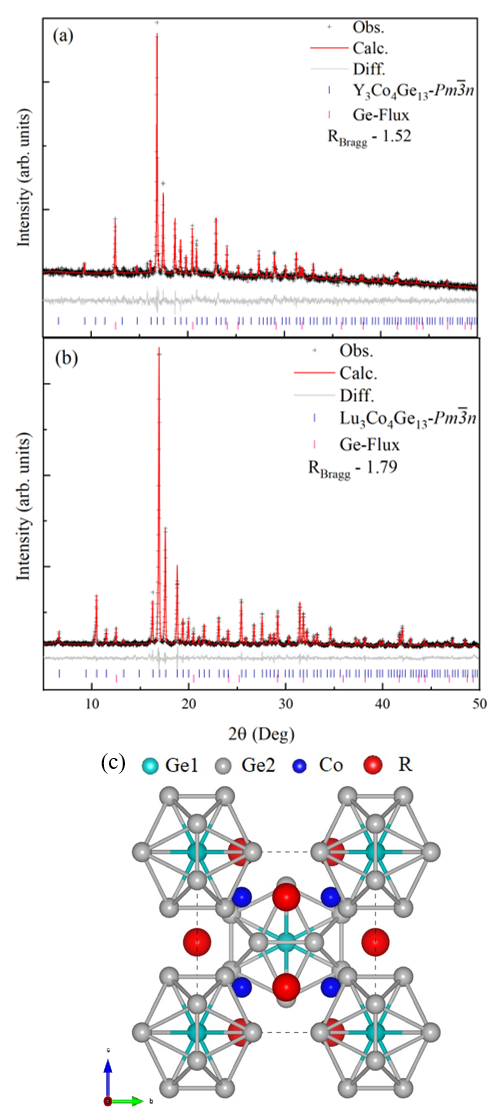}
\caption{Rietveld plot of the room temperature powder XRD pattern for (a) Y$_3$Co$_4$Ge$_{13}$ and (b) Lu$_3$Co$_4$Ge$_{13}$. The black crosses indicate the observed pattern, while the red line represents the calculated one. The gray line at the bottom displays the difference between the experimental and calculated patterns. The blue and pink vertical bars stand for the Bragg reflections. (c) The crystal structure for R$_3$Co$_4$Ge$_{13}$ series in the \textit{Pm$\overline{3}$n} space group.}
\label{fig:fig1}
\end{figure}

The Rietveld refinement of the powder XRD patterns of Y$_3$Co$_4$Ge$_{13}$ and Lu$_3$Co$_4$Ge$_{13}$ compounds are shown in Figs.~\ref{fig:fig1}a and \ref{fig:fig1}b, respectively. The powder XRD patterns of both compounds were successfully indexed by considering the cubic symmetry and \textit{Pm$\overline{3}$n} space group (Fig.~\ref{fig:fig1}) ~\cite{rai2015superconductivity}. The obtained lattice parameters $8{.}726 (2)$~\AA~for Y$_3$Co$_4$Ge$_{13}$ and $8{.}672 (2)$~\AA~ for Lu$_3$Co$_4$Ge$_{13}$, from Rietveld refinement, are consistent with the literature \cite{venturini1985nouvelles}. No spurious phases were detectable other than a small amount of Ge ($<$2$\%$), which is common in flux-grown samples and results from tiny droplets of flux attached to the crystals. Such droplets are carefully removed from the samples used in electrical transport measurements by polishing the slabs before adding electrical contacts.

It is well established that the Remeika phase compounds feature a cage-like structure. Our Rietveld analysis confirms that the Ge atoms in this cage structure have two positions, Ge1 ($2a$) and Ge2 ($24k$), whose occupations were treated as free parameters in the refinement. 
In the literature, it is reported that the thermal vibration at the Ge2 site shows a large prolate thermal ellipsoid \cite{rai2015superconductivity}, which requires to model it as two positionally disordered sites, such as Ge2A, and Ge2B. 
Interestingly, it is also known in these R$_3$Co$_4$Ge$_{13}$ compounds that the interatomic distance between Ge1 and Ge2A/Ge2B sites is much longer than the Ge-Ge bond lengths \cite{rai2015superconductivity}.
Such a large interatomic distance is responsible for an enhanced atomic displacement parameter (ADP) for the Ge1 site, as we can observe from the $B_{eq}$ values refined from our diffraction data (see Table~\ref{table-infoR}), which allows rattling inside the cage. The ADP is determined by crystal structure refinement principally embodying atomic thermal vibration, but it may also include effects of disorder, partial site occupancy, and electronic interactions \cite{chakoumakos1997systematics}.
Rai et al. \cite{rai2015superconductivity} established an empirical relationship between electrical conduction and crystallographic defects and disorder, quantified by the ADP ratio of the transition metals located at the Wyckoff position $8e$ and the Ge atom located at the $2a$ sites. 
In that analysis, it was found that the ADP ratio is less than 3 for metallic compounds, whereas for semiconductor-like behavior, the ADP ratio is around 4.5. 
Following a similar analysis, the ADP ratios for the compounds Y$_3$Co$_4$Ge$_{13}$ and Lu$_3$Co$_4$Ge$_{13}$ were obtained using the $B_{eq}$ parameters from Table~\ref{table-infoR}, ADP ratio = $B_{eq} (2a$ site atom)/$B_{eq}$(transition metal atom). 
For the Y$_3$Co$_4$Ge$_{13}$ and Lu$_3$Co$_4$Ge$_{13}$ compounds, the ADP ratios are 3.21 and 2.97, respectively. This observation suggests that the high structural disorder in Y$_3$Co$_4$Ge$_{13}$ and Lu$_3$Co$_4$Ge$_{13}$ plays a significant role in their electrical transport. Therefore, we have tested several models that describe electron conduction in strongly disordered systems to shed light on the mechanism(s) of electrical transport in Y$_3$Co$_4$Ge$_{13}$ and Lu$_3$Co$_4$Ge$_{13}$.

\begin{table}[htbp]
\caption{Rietveld refined atomic positions, $B_{eq}$ values, and ocupations for R$_3$Co$_4$Ge$_{13}$ (R = Y and Lu).}
\label{table-infoR}
\scriptsize
\begin{center}
\begin{tabular}{c| c| c| c| c |c| c}
\hline
Compound & Wyckoff position & x & y & z & $B_{eq}$ & Occ\\
\hline
Y$_3$Co$_4$Ge$_{13}$&  &  &   &  &  & \\
Y & $6d$ & $\frac{1}{4}$  & 0 & $\frac{1}{2}$ & 1.94 (4) & 1\\
Co & $8e$ & $\frac{1}{4}$ & $\frac{1}{4}$ & $\frac{1}{4}$ & 1.4 (4) & 1\\
Ge1 & $2a$ & 0 & 0 & 0 & {4.5 (7)} & 1\\
Ge2A & $24k$ & 0 & 0 & 1{.}1431 & 1.6 (5) & 0.5\\
Ge2B & $24k$ & 0 & 0 & 0{.}3332  & 0.5 (4) & 0.5\\

Lu$_3$Co$_4$Ge$_{13}$ &  &  &   &  &  & \\
Lu & $6d$ & $\frac{1}{4}$  & 0 & $\frac{1}{2}$ & 1{.}00 (5) & 1\\
Co & $8e$ & $\frac{1}{4}$ & $\frac{1}{4}$ & $\frac{1}{4}$ & 0{.}88 (1) & 1\\
Ge1 & $2a$ & 0 & 0 & 0 & 2{.}62 (2) & 1\\
Ge2A & $24k$ & 0 & 0.2704 & 0{.}1431 & 1.07 (9) & 0.375 \\
Ge2B & $24k$ & 0 & 0.3251 & 0{.}159  & 1.07 (9) & 0.624 \\
\hline
\end{tabular}
\end{center}
\end{table}

\begin{figure}[h]
 \centering
\includegraphics[width=0.6\textwidth]{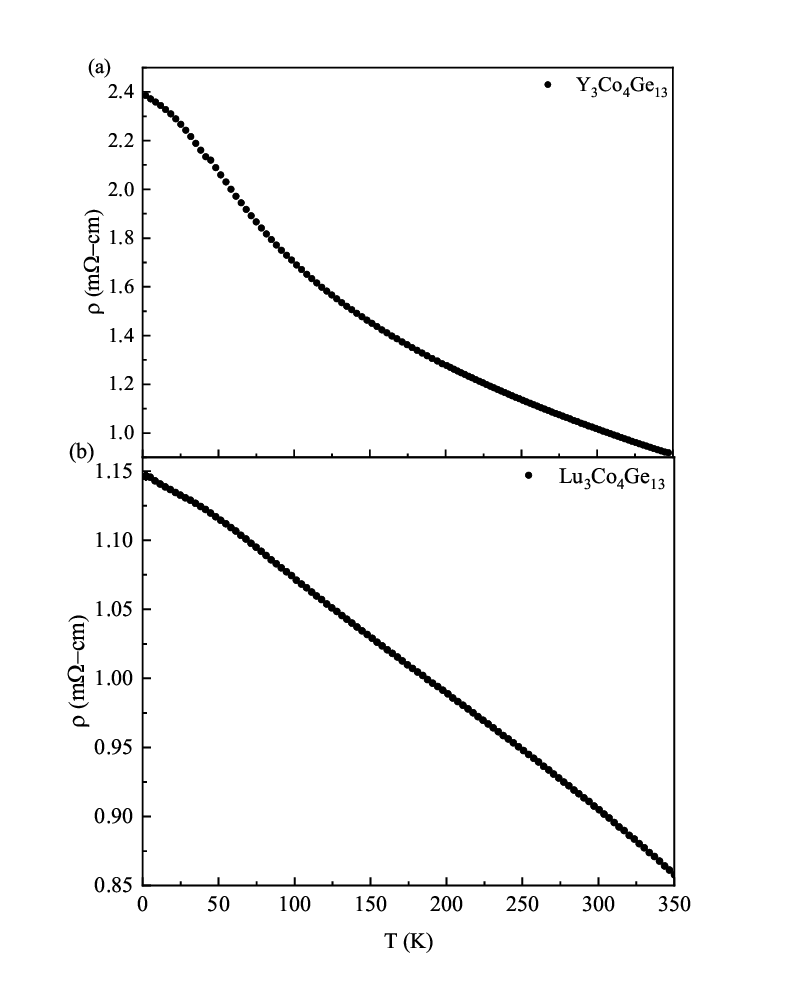}
\caption{Temperature dependence of the electrical resistivity for (a) Y$_3$Co$_4$Ge$_{13}$ and (b) Lu$_3$Co$_4$Ge$_{13}$.}
\label{fig:fig2}
\end{figure}

The temperature-dependent electrical resistivity of the Y$_3$Co$_4$Ge$_{13}$ and Lu$_3$Co$_4$Ge$_{13}$ slabs is shown in Fig.~\ref{fig:fig2}(a) and (b), respectively. 
In contrast to the DFT-based calculations on the density of states (DOS) reported recently \cite{rai2015superconductivity}, the experimental temperature dependence of resistivity for both compounds evidences non-metallic behavior with a negative temperature coefficient \cite{rai2015superconductivity}, also showing a quasi-linear regime of electrical resistivity above $T\approx$ 250~K. 
Below this temperature, for Y$_3$Co$_4$Ge$_{13}$ the $\rho(T)$ rises away from linearity until a gradual leveling-off trend takes place below $\approx$ 25~K. For Lu$_3$Co$_4$Ge$_{13}$, $\rho(T)$ shows only slight changes in slope and a slight tendency of leveling-off towards low temperature. 
With the notable exception of the mixed-valent Yb member \cite{dias2025single}, these general trends of $\rho(T)$ are ubiquitous in the 3-4-13 germanide family \cite{rai2015superconductivity,dias2025single,ghosh1993resistivity}.

\begin{figure}[t]
 \centering
\includegraphics[width=1\textwidth]{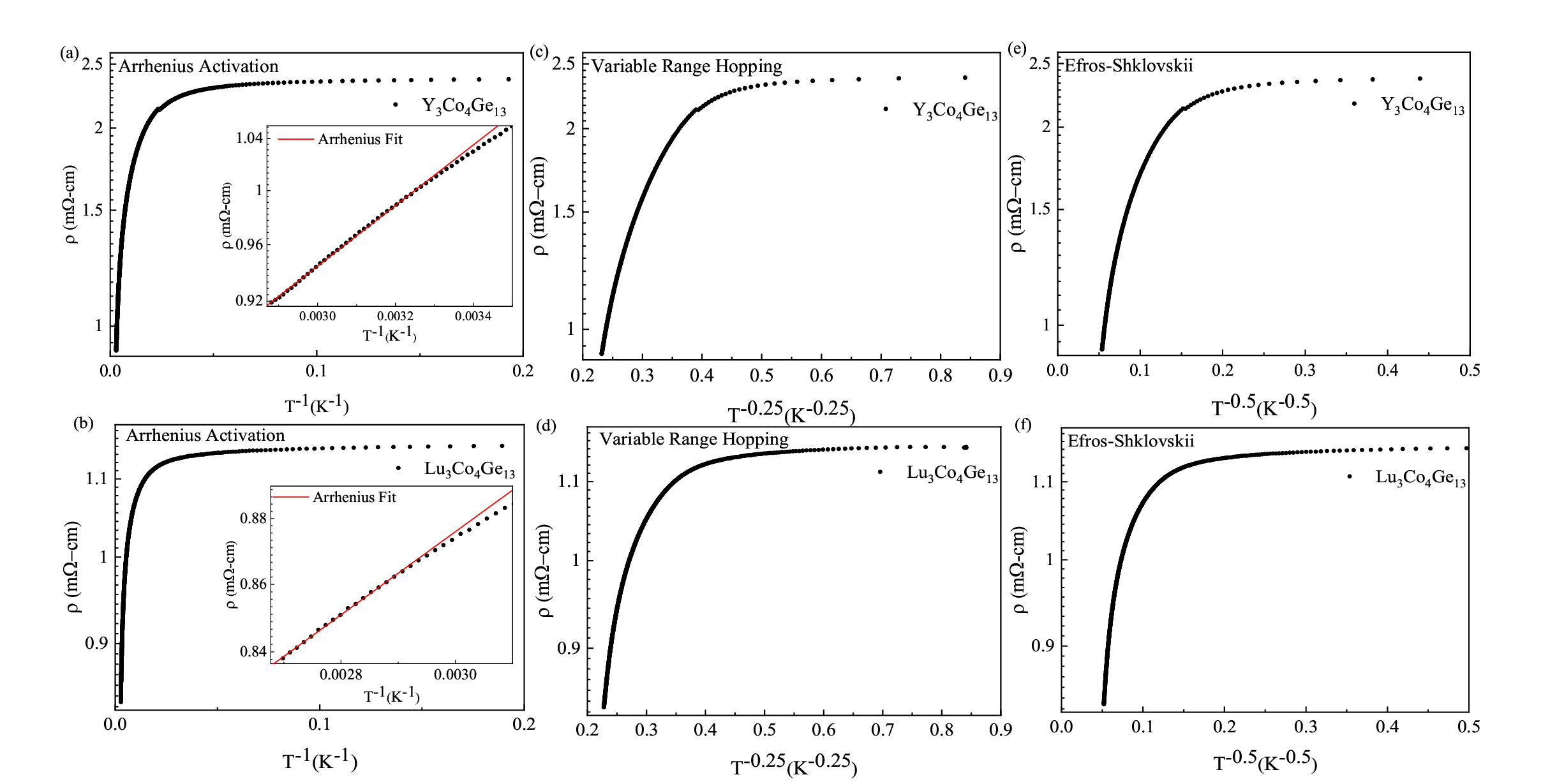}
\caption{Temperature dependence of the electrical resistivity for  Y$_3$Co$_4$Ge$_{13}$ and Lu$_3$Co$_4$Ge$_{13}$. (a) and (b) shows the resistivity as a function of $1/T$, the insets show Arrhenius fit. (c) and (d) The resistivity as a function of $T^{-1/4}$ with VRH model. (e) and (f) The resistivity as a function of $T^{-1/2}$ with Efros-Shlovskii model.}
\label{fig:fig3}
\end{figure}

For typical semiconductors, the temperature dependence of the electrical resistivity can be explained by the Arrhenius law, where electrical conduction occurs due to the thermal activation of charge carriers across the energy gap. 
$\rho(T)$ is described through the exponential temperature dependence, $\rho = \rho_0exp\left({-E_g}/{k_BT}\right)$ \cite{ashcroft2011fisica}, where $\rho_0$ is the resistivity at $T = 0$, $E_g$ is energy gap, and $k_B$ is the Boltzmann constant.
Therefore, to verify whether such a thermal activation mechanism is at play in these samples, we plot $ln(\rho)$ as a function of $1/T$, and the slope of this Arrhenius plot provides the magnitude of the energy gap. 
Fig.~\ref{fig:fig3}(a) and Fig.~\ref{fig:fig3}(b) make it clear that our samples do not follow the pure Arrhenius behavior in an extended temperature range below 300~K. This observation is consistent with the fact that the electrical resistivity in both compounds changes only within a relatively narrower range over the entire temperature interval than what is expected in a typical semiconductor, such as Ge and Si \cite{BULLIS1968639}.
However, linear behavior in both graphs is observed in a rather narrow temperature range above 300~K. 
The extracted energy gaps in that range for Y$_3$Co$_4$Ge$_{13}$ and Lu$_3$Co$_4$Ge$_{13}$ are $E_g/k_B=99.14$~K ($0.008$~eV) and $E_g/k_B=64.06$~K ($0.005$~eV), respectively. 
These values are comparable to the energy gap observed in Pr$_3$Ir$_4$Ge$_{13}$ \cite{kamadurai2021semiconducting} and almost two orders of magnitude smaller than those of the conventional Ge semiconductor at room temperature ($0.67$~eV) \cite{ashcroft2011fisica}. 

In a strongly disordered system, one of the most common mechanisms of electrical conduction is variable range hopping (VRH). 
In this model, electrical conduction at low temperatures is achieved by charge carriers hopping between localized states with characteristic thermal energy \cite{Shklovskii1984}. 
The characteristic temperature dependence of electrical resistivity in the VRH model is defined as $\rho(T) = \rho_0exp\left[{T_0}/{T}\right]^{\beta}$, where $\rho_0$ is the resistivity at $T=0$, $T_0$ a characteristic temperature, and $\beta$ an exponent that takes on values such as $\beta = 1/4$ for the Mott VRH, or $\beta = 1/2$ for the Efros–Shklovskii (ES) VRH \cite{Shklovskii1984} if the electrical conduction also incorporates the Coulomb interaction between the localized electrons. 
Figs.~\ref{fig:fig3}(c)-(f) show that neither of these mechanisms are sufficient to explain $\rho(T)$ of our samples in an extended temperature range. 

Thus, we proceed to analyze the electrical resistivity in the entire measured temperature range of 2-350~K by considering a model that has recently been adopted to explain $\rho(T)$ in disordered half-Heusler intermetallic compounds \cite{gnida2021origin, koushik2024unveiling, gofryk2007magnetic}. It is based on two parallel conductivity channels, namely metallic-like ($\sigma_m$): 

\begin{equation}
    \sigma_m = (\rho_0 + aT)^{-1},
    \label{sigma_m}
\end{equation}

\noindent and semiconducting-like ($\sigma_s$):

\begin{equation}
  \sigma_s = \sigma_a exp\left(\frac{-E_g}{2k_BT}\right).
   \label{sigma_s}
\end{equation}

\noindent With the respective contribution to $\rho(T)$ of both channels represented by \cite{gnida2021origin}

\begin{equation}
    \rho = (\sigma_m+\sigma_s)^{-1}.
    \label{eq:rho}
\end{equation}

In these equations $\rho_0$ is the residual resistivity, $aT$ is a simple approximation of the electron-phonon scattering contribution, $E_g$ is the semiconducting energy gap, and $k_B$ is the Boltzmann constant.

\begin{figure}[h]
 \centering
\includegraphics[width=0.45\textwidth]{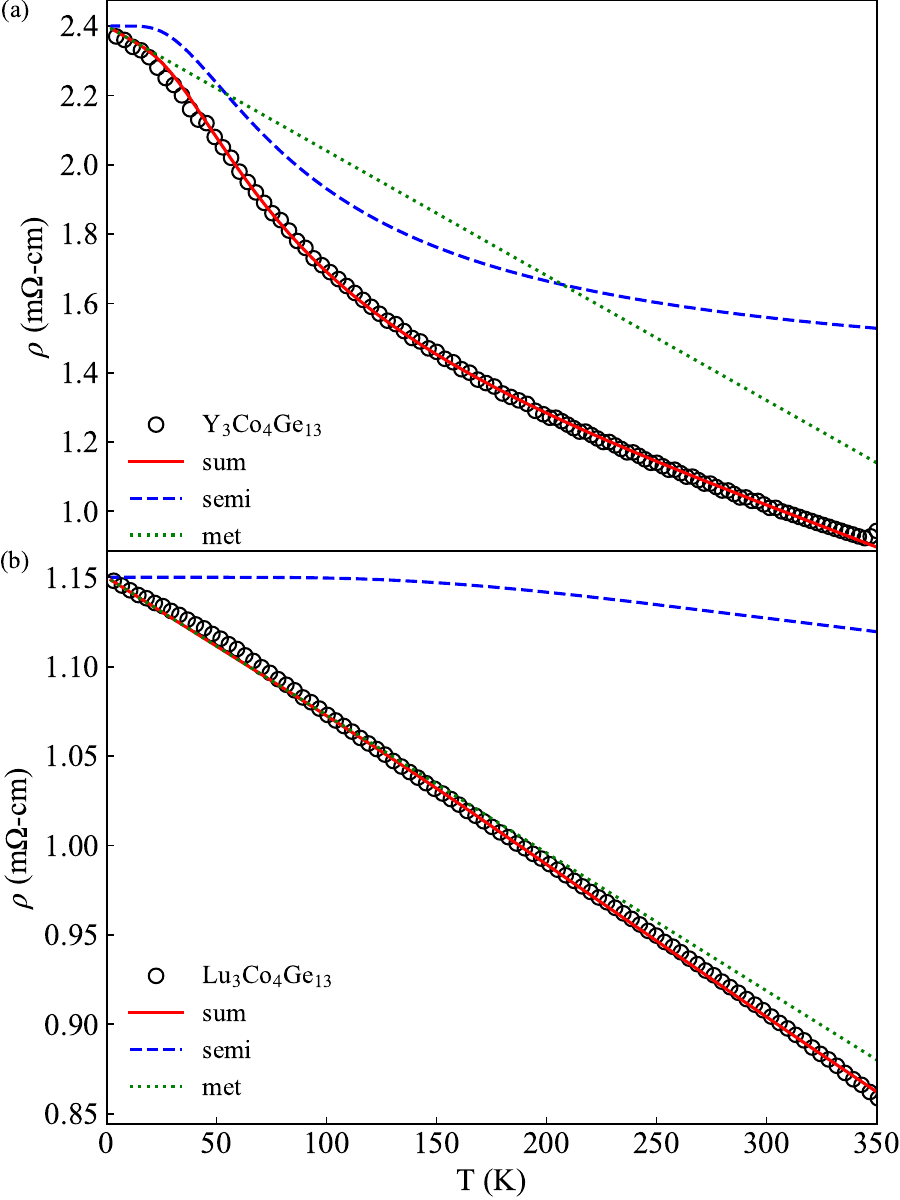}
\caption{Temperature dependence of the electrical resistivity of (a) Y$_3$Co$_4$Ge$_{13}$ and (b) Lu$_3$Co$_4$Ge$_{13}$. Solid lines represent the fitting of Eq.~\ref{eq:rho}, the dashed line (blue) is the simulated semiconductor channel with Eq.~\ref{sigma_m}, and the dotted line (green) is simulated with Eq.~\ref{sigma_s} the metallic channel.}
\label{fig:fig4}
\end{figure}

The fittings of this model are shown in Fig.~\ref{fig:fig4} and the fitted values obtained for Y$_3$Co$_4$Ge$_{13}$ (Fig.~\ref{fig:fig4}(a)) and Lu$_3$Co$_4$Ge$_{13}$ (Fig.~\ref{fig:fig4}(b)) are listed in Table~\ref{table_results_eq3}.

\begin{table}[htbp]
\begin{center}
\caption{Fitting parameters from Eq.~\ref{eq:rho}}
\label{table_results_eq3}
\begin{tabular}{l | c | c | c | c}
\hline
Compound & $\rho_0$ (m$\Omega$-cm) & $a$ (\text{m}$\Omega$-cm~K$^{-1}$) & $\sigma_a$ (m$\Omega$-cm)$^{-1}$ & $E_g$ (meV)\\
\hline
Y$_3$Co$_4$Ge$_{13}$ & 2.4  & $-3.6 (3) \times10^{-6}$  & 0.336 (3) & 20(4)\\
Lu$_3$Co$_4$Ge$_{13}$ & 1.15 & $-0.771 (3)\times10^{-6}$ & 0.136(3) & 106(3)\\
\hline
\end{tabular}
\end{center}
\end{table}

Clearly, $\rho(T)$ for both samples can be much better described using the two-channel conduction model, albeit the fitted curves do show some deviations at low temperatures, which could be related to some additional, unaccounted for scattering mechanism(s). 
For the Y compound (Fig.~\ref{fig:fig4}(a)), both channels contribute significantly to $\rho(T)$, with the majority response being semiconducting close to 80~K. 
For the Lu compound (Fig.~\ref{fig:fig4}(b)), the fit deviates somewhat below $\approx 80$~K and the metallic contribution is dominant over the entire temperature range. 
At very low temperatures ($\approx$10~K), for both samples, the semiconducting channel does not contribute significantly and the electronic transport is dominated by the metallic channel.

Interestingly, the coefficient of the linear-$T$ term in the two-channel conduction model is found to be negative for both samples ($a<0$). 
Although its magnitude $|a|$ is small compared to those observed in half-Heusler compounds, its negative value cannot be explained using conventional electron-phonon scattering mechanisms. 
Such negative coefficients in the linear-$T$ term have been observed previously in several strongly disordered metals, including the irradiation-damaged Nb$_3$Ge and V$_3$Si compounds \cite{PhysRevB.15.2570}. 

In his seminal work \cite{mooij1973electrical}, Mooij concluded that a negative coefficient in the linear-$T$ term is ubiquitous in strongly disordered system, and that it is a rule rather than an exception for metals with resistivity larger than $150~\mu\Omega-$cm.
The unusual electrical transport properties of 3-4-13 germanides have often been associated with the presence of structural disorder due to their unique crystal structure as discussed above \cite{strydom2007thermal, prakash2013superconductivity, rai2015superconductivity, afzal2024superconductivity, prakash2015multiband, kumar2018low}, although their temperature variation has not been investigated in detail. 
Recent DFT calculations support such claims, since they predict that the 3-4-13 germanides should be poor metals \cite{rai2015superconductivity}.

In literature, the observed negative temperature coefficient of electrical resistivity in disordered metallic systems has been described by the so-called ``structural Kondo effect'' \cite{tsuei1978anomalous, PhysRevB.22.2665}. 
The well-known traditional Kondo effect is related to the exchange interaction between magnetic impurities and conduction electrons, which leads to the famous ``Kondo minima'' in resistivity upon cooling, followed by a logarithmic ($\sim$ln$T$) increase regime at lower temperatures \cite{10.1143/PTP.32.37}. 
Extending the concept to a similar scenario, the structural Kondo effect is the non-magnetic analog of the traditional Kondo effect, where the conduction electrons couple to a suitable low-energy degree of freedom made available by structural disorder \cite{tsuei1978anomalous}. 
In the present case, the structural disorder signifies structural indeterminency due to the large atomic displacement at Ge1 and Ge2 sites, as well as the rattling of Ge atom inside an oversized cage.

\begin{figure}[htpb]
 \centering
\includegraphics[width=0.6\textwidth]{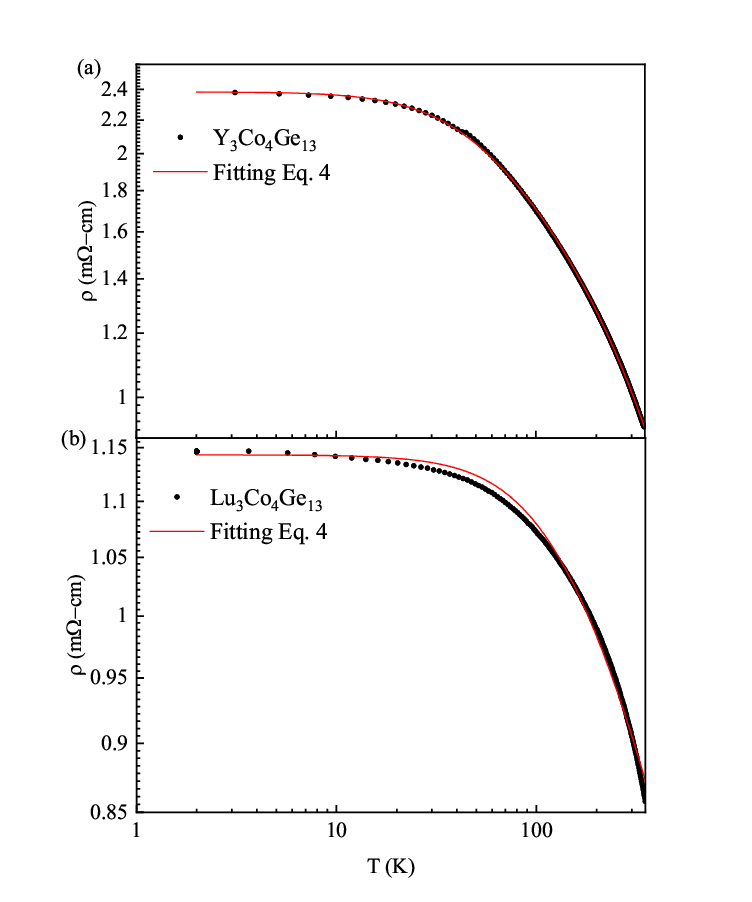}
\caption{Temperature dependence of the electrical resistivity on a logarithmic scale of the (a) Y$_3$Co$_4$Ge$_{13}$ (b) and Lu$_3$Co$_4$Ge$_{13}$. The solid line represents the fitting with Eq.~\ref{Kondo}.}
\label{fig:fig5}
\end{figure}

By employing the ``double-well tunneling model'' Cochrane \textit{et al.} explained the Kondo-like minima, and $\sim$ln$T$, in electrical resistivity at high temperatures in non-magnetic disordered metal systems, using the following expression \cite{PhysRevLett.35.676, tsuei1978anomalous, PhysRevB.22.2665}:

\begin{equation}
    \rho(T) = A+C\ln(T^2+T_\Delta^2),
    \label{Kondo}
\end{equation}

\noindent where $A$ is a constant, $C$ is a measure of the concentration of the effective tunneling configurations, and $T_\Delta$ marks the beginning of a gradual transition from high-temperature state to low-temperature state.

Equation \ref{Kondo} reasonably fitted the experimental data in the entire measured temperature range of both our samples, as demonstrated in Fig.\ref{fig:fig5}(a) and (b) on a logarithmic scale. 
For Y$_3$Co$_4$Ge$_{13}$ (Fig.~\ref{fig:fig5}(a)) the obtained fitting parameters are, $A=4.81$~m$\Omega-$cm, $C=-0.33$~m$\Omega-$cm, and $T_\Delta=38.5$~K, and for Lu$_3$Co$_4$Ge$_{13}$ (Fig.~\ref{fig:fig5}(b)), $A=2.28$~m$\Omega-$cm, $C=-0.11$~m$\Omega-$cm and $T_\Delta=118.9$~K. 
The value of the parameter $C$, obtained for both samples, is two orders of magnitude larger than the values found in Refs. \cite{tsuei1978anomalous, PhysRevB.22.2665} for amorphous and disordered transition-metal alloy systems. 
This may be due to the significantly larger resistivity values in both samples presented here than the ones discussed in Refs. \cite{tsuei1978anomalous, PhysRevB.22.2665}. 
Moreover, the obtained values for the parameter $T_{\Delta}$, for both samples, are found to be consistent with Refs. \cite{tsuei1978anomalous, PhysRevB.22.2665}. 

As pointed out in Ref. \cite{tsuei1978anomalous}, the double-well potential mechanism is not essential for Eq.~\ref{Kondo}, as originally proposed by Cochrane \textit{et al.} \cite{PhysRevLett.35.676}. 
However, any localized internal degree of freedom (such as local phonons) that gives rise to excitations can lead to a negative temperature coefficient of resistivity through this structural Kondo mechanism. 
The prevalent Einstein phonon modes below 100 K, reported recently in both samples \cite{dias2025single}, seem to be an appropriate internal degree of freedom to which the conduction electron may couple in the structural Kondo mechanism. 
It is worth mentioning that the double-well potential model proposed by Cochrane \textit{et al.} has also been used to explain superconductivity in strongly disordered or amorphous alloy systems \cite{RIESS1980334}. 
Therefore, it is worth investigating the superconductivity in intrinsically disordered 3-4-13 germanides using this model in future work \cite{rai2015superconductivity}.

\section{Summary and Conclusion}
\label{sec4}

The temperature dependence of electrical resistivity is investigated in the Remeika phases Y$_3$Co$_4$Ge$_{13}$, and Lu$_3$Co$_4$Ge$_{13}$ single crystals. 
The samples were prepared using the self-flux method and carefully characterized via powder XRD and EDXS. 
Both samples exhibit semiconducting-like behavior, with resistivity increasing as temperature decreases. 
This behavior contradicts density of states (DOS) calculations, which predict metallic behavior in these compounds. 
Thus, the observed semiconducting-like transport is most likely associated with strong structural disorder arising from their unique crystal structure, and low carrier density. 
Despite the semiconducting-like resistivity behavior, its temperature dependence does not follow the conventional Arrhenius thermal activation mechanism below 350 K, nor does it align with variable-range hopping models commonly observed in disordered metals. 
However, the entire temperature dependence of resistivity can be reasonably described using a two-channel conduction model, where electron transport occurs through both semiconducting and metallic channels in parallel. 
Via carefully analyzing the experimental data, we observed that the semiconducting channel dominates at high temperatures, while the metallic channel becomes the primary conduction path below 10 K. 
Additionally, we observe a negative temperature coefficient of resistivity in the metallic conduction channel. 
Although the exact origin of this behavior is not yet fully understood, we argue that it can be ascribed to strong structural disorder, by invoking the structural Kondo mechanism. 
We propose that the significant disorder affecting electron conduction in these samples may also play a role in the unconventional superconductivity observed in the Remeika phases 3-4-13 germanides.  

\section*{Acknowledgments}
\label{app5}

The authors acknowledge financial support of the INCT project Advanced Quantum Materials, involving the Brazilian agencies CNPq (Proc. 408766/2024-7, 142585/ 2020-3 and 306827/2023-9), FAPESP (Grant \#s 2017/20989-8, 2017/10581-1 and 2023/09820-2), and CAPES.
The authors thank the Multiuser Central Facilities (UFABC) for the experimental support. This work also used facilities of the National Nanotechnology Laboratory (LNNano), part of the Brazilian Centre for Research in Energy and Materials (CNPEM), a private non-profit organization under the supervision of the Brazilian Ministry for Science, Technology, and Innovations (MCTI). The Thin Films and Electrical Measurements Laboratory and Nanomaterials Division staff are acknowledged for the assistance during the experiments (Proposal \# 20240454).

\bibliographystyle{elsarticle-num.bst}
\bibliography{Bibliography}

@article{remeika1980new,
  title={A new family of ternary intermetallic superconducting/magnetic stannides},
  author={Remeika, JP and Espinosa, GP and Cooper, AS and Barz, H and Rowell, JM and McWhan, DB and Vandenberg, JM and Moncton, DE and Fisk, Z and Woolf, LD and others},
  journal={Solid State Communications},
  volume={34},
  number={12},
  pages={923--926},
  year={1980},
  publisher={Elsevier}
}

@article{oswald2017proof,
  title={The proof is in the powder: revealing structural peculiarities in the {Yb$_3$Rh$_4$Sn$_{13}$} structure type},
  author={Oswald, Iain WH and Rai, Binod K and McCandless, Gregory T and Morosan, Emilia and Chan, Julia Y},
  journal={CrystEngComm},
  volume={19},
  number={25},
  pages={3381--3391},
  year={2017},
  publisher={Royal Society of Chemistry}
}

@article{BULLIS1968639,
title = {Temperature coefficient of resistivity of silicon and germanium near room temperature},
journal = {Solid-State Electronics},
volume = {11},
number = {7},
pages = {639-646},
year = {1968},
issn = {0038-1101},
author = {W.M. Bullis and F.H. Brewer and C.D. Kolstad and L.J. Swartzendruber},
}

@incollection{GUMENIUK201843,
title = {Chapter 304 - Structural and Physical Properties of Remeika Phases},
editor = {Jean-Claude G. Bünzli and Vitalij K. Pecharsky},
series = {Handbook on the Physics and Chemistry of Rare Earths},
publisher = {Elsevier},
volume = {54},
pages = {43-143},
year = {2018},
booktitle = {Including Actinides},
issn = {0168-1273},
author = {Roman Gumeniuk},
}

@incollection{UHER2001139,
title = {Chapter 5 Skutterudites: Prospective novel thermoelectrics},
editor = {Terry M. Tritt},
series = {Semiconductors and Semimetals},
publisher = {Elsevier},
volume = {69},
pages = {139-253},
year = {2001},
booktitle = {Recent Trends in Thermoelectric Materials Research I},
issn = {0080-8784},
author = {Ctirad Uher},
}

@article{prakash2013superconductivity,
  title={Superconductivity in a low carrier density system: {A} single crystal study of cubic {Y$_3$Ru$_4$Ge$_{13}$}},
  author={Prakash, Om and Thamizhavel, A and Nigam, AK and Ramakrishnan, S},
  journal={Physica C: Superconductivity},
  volume={492},
  pages={90--95},
  year={2013},
  publisher={Elsevier}
}

@article{rai2015superconductivity,
  title={Superconductivity in Single Crystals of {Lu$_3$T$_4$Ge$_{13-x}$} ({T= Co, Rh, Os}) and {Y$_3$T$_4$Ge$_{13-x}$} ({T= Ir, Rh, Os})},
  author={Rai, Binod K and Oswald, Iain WH and Wang, Jiakui K and McCandless, Gregory T and Chan, Julia Y and Morosan, E},
  journal={Chemistry of Materials},
  volume={27},
  number={7},
  pages={2488--2494},
  year={2015},
  publisher={ACS Publications}
}

@inproceedings{prakash2014superconductivity,
  title={Superconductivity in {Y$_3$Ru$_4$Ge$_{13-x}$} and {Lu$_3$Os$_4$Ge$_{13-x}$}: {A} comparative study},
  author={Prakash, Om and Thamizhavel, A and Ramakrishnan, S},
  booktitle={Journal of Physics: Conference Series},
  volume={568},
  number={2},
  pages={022039},
  year={2014},
  organization={IOP Publishing}
}

@article{rai2016intermediate,
  title={Intermediate valence to heavy fermion through a quantum phase transition in {Yb$_3$(Rh$_{1-x}$T$_x$)$_4$Ge$_{13}$} ({T= Co, Ir}) single crystals},
  author={Rai, Binod K and Oswald, Iain WH and Chan, Julia Y and Morosan, E},
  journal={Physical Review B},
  volume={93},
  number={3},
  pages={035101},
  year={2016},
  publisher={APS}
}

@article{nolas1999skutterudites,
  title={Skutterudites: {A} phonon-glass-electron crystal approach to advanced thermoelectric energy conversion applications},
  author={Nolas, GS and Morelli, DT and Tritt, Terry M},
  journal={Annual Review of Materials Science},
  volume={29},
  number={1},
  pages={89--116},
  year={1999},
  publisher={Annual Reviews 4139 El Camino Way, PO Box 10139, Palo Alto, CA 94303-0139, USA}
}

@article{rull2015skutterudites,
  title={Skutterudites as thermoelectric materials: revisited},
  author={Rull-Bravo, Marta and Moure, Alberto and Fern{\'a}ndez, JF and Mart{\'\i}n-Gonz{\'a}lez, Marisol},
  journal={Rsc Advances},
  volume={5},
  number={52},
  pages={41653--41667},
  year={2015},
  publisher={Royal Society of Chemistry}
}

@article{chakoumakos1997systematics,
  title={Systematics of atomic displacement parameters in perovskite oxides},
  author={Chakoumakos, BC},
  journal={Physica B: Condensed Matter},
  volume={241},
  pages={361--363},
  year={1997},
  publisher={Elsevier}
}

@article{coelho2018topas,
  title={TOPAS and TOPAS-Academic: an optimization program integrating computer algebra and crystallographic objects written in C++},
  author={Coelho, Alan A},
  journal={Applied Crystallography},
  volume={51},
  number={1},
  pages={210--218},
  year={2018},
  publisher={International Union of Crystallography}
}

@article{kataria2023time,
  title={Time-reversal symmetry breaking in the superconducting low carrier density quasiskutterudite {Lu$_3$Os$_4$Ge$_{13}$}},
  author={Kataria, A and Verezhak, JAT and Prakash, O and Kushwaha, RK and Thamizhavel, A and Ramakrishnan, S and Scheurer, MS and Hillier, AD and Singh, RP},
  journal={Physical Review B},
  volume={107},
  number={10},
  pages={L100506},
  year={2023},
  publisher={APS}
}

@article{gnida2021origin,
  title={Origin of the negative temperature coefficient of resistivity in the half-{H}eusler antimonides {LuNiSb} and {YPdSb}},
  author={Gnida, Daniel and Ciesielski, Kamil and Kaczorowski, Dariusz},
  journal={Physical Review B},
  volume={103},
  number={17},
  pages={174206},
  year={2021},
  publisher={APS}
}

@article{kamadurai2021semiconducting,
  title={Semiconducting behaviour in the {R}emeika phase: {Pr$_3$Ir$_4$Ge$_{13}$}},
  author={Kamadurai, Ramesh Kumar and Ogunbunmi, Michael O and Nair, Harikrishnan S and Strydom, Andr{\'e} M},
  journal={Journal of Alloys and Compounds},
  volume={872},
  pages={159481},
  year={2021},
  publisher={Elsevier}
}

@article{koushik2024unveiling,
  title={Unveiling the structure-property relationship in a disordered inverse {H}eusler alloy {Ti$_2$MnAl}},
  author={Koushik, P and Mukherjee, K},
  journal={Physics Letters A},
  pages={129628},
  year={2024},
  publisher={Elsevier}
}

@article{mooij1973electrical,
  title={Electrical conduction in concentrated disordered transition metal alloys},
  author={Mooij, JH},
  journal={physica status solidi (a)},
  volume={17},
  number={2},
  pages={521--530},
  year={1973},
  publisher={Wiley Online Library}
}

@article{prakash2015multiband,
  title={Multiband superconductivity in {Lu$_3$Os$_4$Ge$_{13}$}},
  author={Prakash, Om and Thamizhavel, A and Ramakrishnan, S},
  journal={Superconductor Science and Technology},
  volume={28},
  number={11},
  pages={115012},
  year={2015},
  publisher={IOP Publishing}
}

@article{afzal2024superconductivity,
  title={Superconductivity in chiral cubic {Y$_3$Rh$_4$Ge$_{13}$}},
  author={Afzal, Md Asif and Higashinaka, Ryuji and Iwasa, Kazuaki and Ahmed, Nazir and Tsubota, Ryo and Nakamura, Naoki and Matsuda, Tatsuma D and Aoki, Yuji},
  journal={Journal of Alloys and Compounds},
  volume={978},
  pages={172914},
  year={2024},
  publisher={Elsevier}
}

@article{gofryk2007magnetic,
  title={Magnetic, transport, and thermal properties of the half-{H}eusler compounds {ErPdSb} and {YPdSb}},
  author={Gofryk, Krzysztof and Kaczorowski, D and Plackowski, T and Mucha, J and Leithe-Jasper, A and Schnelle, W and Grin, Yu},
  journal={Physical Review B—Condensed Matter and Materials Physics},
  volume={75},
  number={22},
  pages={224426},
  year={2007},
  publisher={APS}
}

@article{venturini1985nouvelles,
  title={De nouvelles séries de germaniures, isotypes de {Yb$_3$Rh$_4$Sn$_{13}$ et BaNiSn$_3$}, dans les systèmes ternaires {TR$-$T$-$Ge où TR} est un element des terres rares et {T$\equiv$ Co, Rh, Ir, Ru, Os}},
  author={Venturini, G and M{\'e}ot-Meyer, M and Malaman, B and Roques, B},
  journal={Journal of the Less Common Metals},
  volume={113},
  number={2},
  pages={197--204},
  year={1985},
  publisher={Elsevier}
}

@inproceedings{kumar2018low,
  title={Low carrier semiconductor like behavior in {Lu$_3$Ir$_4$Ge$_{13}$} single crystal},
  author={Kumar, Anil and Matteppanavar, Shidaling and Thamizhavel, A and Ramakrishnan, S},
  booktitle={AIP Conference Proceedings},
  volume={1942},
  number={1},
  year={2018},
  organization={AIP Publishing}
}

@article{strydom2007thermal,
  title={Thermal and transport properties of the cubic semimetal {Y$_3$Ir$_4$Ge$_{13}$}: on the metallic border of thermoelectric merit},
  author={Strydom, AM},
  journal={Journal of Physics: Condensed Matter},
  volume={19},
  number={38},
  pages={386205},
  year={2007},
  publisher={IOP Publishing}
}

@book{ashcroft2011fisica,
  title={F{\'\i}sica do estado s{\'o}lido},
  author={Ashcroft, Neil W and Mermin, N David},
  year={2011},
  publisher={Cengage Learning}
}

@article{ghosh1993resistivity,
  title={Resistivity and magnetic-susceptibility studies in the {R$_3$Ru$_4$Ge$_{13}$} ({R= Nd, Dy, Ho, Er, Yb, Lu, and Y}) system},
  author={Ghosh, K and Ramakrishnan, S and Chandra, Girish},
  journal={Physical Review B},
  volume={48},
  number={14},
  pages={10435},
  year={1993},
  publisher={APS}
}

@article{dias2025single,
  title={Single crystal growth and characterization of the intermetallic cage compounds {R$_3$Co$_4$Ge$_{13}$ (R$=$ Y, Gd$-$Lu)}},
  author={Dias, Juliana Gonçalves and Santos, Pedro Caetano Sabino and Vasques, Gustavo Gomes and Dutra, Mateus Souza and Mendonça-Ferreira, Leticie and Avila, Marcos A},
  journal={Journal of {A}lloys and {C}ompounds},
  volume = {1012},
  pages={178479},
  year={2025},
  publisher={Elsevier}
}

@article{feig2021valence,
  title={Valence fluctuations in the {3D$+$3} modulated {Yb$_{3}$Co$_{4}$Ge$_{13}$ R}emeika phase},
  author={Feig, Manuel and Akselrud, Lev and Motylenko, Mykhaylo and Bobnar, Matej and Wagler, J{\"o}rg and Kvashnina, Kristina O and Levytskyi, Volodymyr and Rafaja, David and Leithe-Jasper, Andreas and Gumeniuk, Roman},
  journal={Dalton Transactions},
  volume={50},
  number={38},
  pages={13580--13590},
  year={2021},
  publisher={Royal Society of Chemistry}
}

@article{tsuei1978anomalous,
  title={Anomalous electrical conduction in disordered and non-crystalline metallic conductors},
  author={Tsuei, CC},
  journal={Solid State Communications},
  volume={27},
  number={7},
  pages={691--695},
  year={1978},
  publisher={Elsevier}
}

@article{PhysRevB.22.2665,
  title = {Negative-temperature coefficients of electrical resistivity in amorphous {La-based} alloys},
  author = {Mueller, R. and Agyeman, K. and Tsuei, C. C.},
  journal = {Phys. Rev. B},
  volume = {22},
  issue = {6},
  pages = {2665--2669},
  year = {1980},
  month = {Sep},
  publisher = {American Physical Society}
}

@article{mott1970conduction,
  title={Conduction in non-crystalline systems: {IV. Anderson} localization in a disordered lattice},
  author={Mott, NF},
  journal={Philosophical Magazine},
  volume={22},
  number={175},
  pages={7--29},
  year={1970},
  publisher={Taylor \& Francis}
}

@article{PhysRevB.15.2570,
  title = {Anomalous electrical resistivity and defects in $A\ensuremath{-}15$ compounds},
  author = {Testardi, L. R. and Poate, J. M. and Levinstein, H. J.},
  journal = {Phys. Rev. B},
  volume = {15},
  issue = {5},
  pages = {2570--2580},
  numpages = {0},
  year = {1977},
  month = {Mar},
  publisher = {American Physical Society}
}

@article{PhysRevLett.35.676,
  title = {Structural Manifestations in Amorphous Alloys: Resistance Minima},
  author = {Cochrane, R. W. and Harris, R. and Str\"om-Olson, J. O. and Zuckermann, M. J.},
  journal = {Phys. Rev. Lett.},
  volume = {35},
  issue = {10},
  pages = {676--679},
  numpages = {0},
  year = {1975},
  month = {Sep},
  publisher = {American Physical Society}
}

@article{RIESS1980334,
title = {Superconductivity pairing induced by two level systems in amorphous metals},
journal = {Physics Letters A},
volume = {79},
number = {4},
pages = {334-336},
year = {1980},
issn = {0375-9601},
author = {J. Riess and R. Maynard}
}

@article{10.1143/PTP.32.37,
    author = {Kondo, Jun},
    title = {Resistance Minimum in Dilute Magnetic Alloys},
    journal = {Progress of Theoretical Physics},
    volume = {32},
    number = {1},
    pages = {37-49},
    year = {1964},
    month = {07}
}

@article{ZHANG2020152272,
title = {Characterization of multiple-filled skutterudites with high thermoelectric performance},
journal = {Journal of Alloys and Compounds},
volume = {814},
pages = {152272},
year = {2020},
issn = {0925-8388},
author = {Shuye Zhang and Sunwu Xu and Hui Gao and Qingshuang Lu and Tiesong Lin and Peng He and Huiyuan Geng}
}

@Inbook{Shklovskii1984,
author="Shklovskii, Boris I.
and Efros, Alex L.",
title="Variable-Range Hopping Conduction",
bookTitle="Electronic Properties of Doped Semiconductors",
year="1984",
publisher="Springer Berlin Heidelberg",
address="Berlin, Heidelberg",
pages="202--227"
}

@article{MOROZKIN2013121,
title = {Thermoelectric properties of {Pr$_3$Rh$_4$Sn$_{13}$-type Yb$_3$Co$_4$Ge$_{13}$ and Yb$_3$Co$_4$Sn$_{13}$} compounds},
journal = {Journal of Alloys and Compounds},
volume = {549},
pages = {121-125},
year = {2013},
issn = {0925-8388},
author = {A.V. Morozkin and V.Yu. Irkhin and V.N. Nikiforov},
}

@article{Ogunbunmi_2020,
year = {2020},
month = {jul},
publisher = {IOP Publishing},
volume = {32},
number = {40},
pages = {405606},
author = {Ogunbunmi, Michael O and Strydom, André M},
title = {Promising thermoelectric properties of heavy-fermion semimetal {Pr$_3$Os$_4$Ge$_{13}$}},
journal = {Journal of Physics: Condensed Matter},
}





\end{document}